\documentclass[pra,amsmath,amssymb,twocolumn,superscriptaddress,showpacs]{revtex4}

\usepackage{amsmath,amssymb}
\usepackage[usenames]{color}
\usepackage{mathbbol}
\usepackage{amssymb}
\usepackage{grffile}
\usepackage[pdftex]{graphicx}
\usepackage{amsmath, amstext, amssymb, amsfonts, amsxtra}
\usepackage{textcomp}
\usepackage{xspace}
\DeclareSymbolFontAlphabet{\amsmathbb}{AMSb}

\newcommand{\aop}{\hat{a}} 
\newcommand{\bop}{\hat{b}} 
\newcommand{\cop}{\hat{c}} 
    
\newcommand{\nop}{\hat{n}} 
\newcommand{\Op}{\hat{O}}   
\newcommand{\adop}{\hat{a}^{\dagger}} 
 
\newcommand{\cdop}{\hat{c}^{\dagger}}

\newcommand{\cur}{\mathcal{J}}

\newcommand{\rhop}{\hat{\rho}}

\newcommand{\Jp}{J^{\perp}}   
\newcommand{\Jpa}{J^{\parallel}}   
 
\newcommand{\Hop}{\hat{H}} 
\newcommand{\Dop}{\mathcal{D}}
\newcommand{\lan}{\left\langle} 
\newcommand{\ran}{\right\rangle} 
 
\newcommand{\im}{{\rm i}}   
\newcommand{\alpop}{\hat{\alpha}} 
\newcommand{\Am}{\bold{A}} 
\newcommand{\Bm}{\bold{B}} 
\newcommand{\Cm}{\bold{C}} 
\newcommand{\Dm}{\bold{D}} 
\newcommand{\Xm}{\bold{X}}
\newcommand{\Pm}{\bold{P}}
\newcommand{\Hm}{\bold{H}} 
\newcommand{\hm}{\bold{h}}
\newcommand{\Sm}{\bold{S}} 
\newcommand{\Um}{\bold{U}} 
\newcommand{\Vm}{\bold{V}}
\newcommand{\Gm}{\bold{\Gamma}}
\newcommand{\Lmp}{\bold{\Lambda}^{+}}
\newcommand{\Lmm}{\bold{\Lambda}^{-}}
\newcommand{\one}{\bold{1}}

\newcommand{\sutd}{Singapore University of Technology and Design, 8 Somapah Road, 487372 Singapore} 

\begin{document}

\title{Geometry of system-bath coupling and gauge fields in bosonic ladders: manipulating currents and driving phase transitions}                     

\author{Chu Guo} 
\affiliation{\sutd}
\author{Dario Poletti}
\affiliation{\sutd}

\begin{abstract} 
Quantum systems in contact with an environment display a rich physics emerging from the interplay between dissipative and Hamiltonian terms. Here we focus on the role of the geometry of the coupling between the system and the baths. In the specific we consider a dissipative boundary driven ladder in presence of a gauge field which can be implemented with ion microtraps arrays. We show that, depending on the geometry, the currents imposed by the baths can be strongly affected by the gauge field resulting in non-equilibrium phase transitions. In different phases both the magnitude of the current and its spatial distribution are significantly different. These findings allow for novel strategies to manipulate and control transport properties in quantum systems. 
\end{abstract}


\maketitle

\section{Introduction} 

Quantum systems in contact with an environment can present a very rich phenomenology due to the interplay between the Hamiltonian and dissipative dynamics. While the Hamiltonian can allow the emergence of quantum phase transitions, the dissipator can drive the system to a particular steady state (or more than one) the properties of which vary depending on the Hamiltonian itself. In these systems non-equilibrium phase transitions may occur \cite{DiehlZoller2008, DiehlZoller2010, DallaTorreAltman2010, LudwigMarquardt2013}. Recent advances on quantum fluids of light \cite{CarusottoCiuti2013}, and on ultracold gases in cavities \cite{BaumannEsslinger2010} have allowed experimental exploration of this physics. 

A particularly important set of open systems is the boundary driven systems in which the set-up is connected at its boundaries to the environment. A typical example is that of a chain of particles connected, at its extremities, to two different baths, thus imposing a flow within the system. This scenario has been studied both theoretically and experimentally for various kinds of physical realizations including spin systems, photonic lattices, ions, metals, semiconductors and Josephson junctions arrays \cite{MichelMahler2006, BenentiPeyrard2015}. Experiments with ion traps promise to be an ideal set-up to simulate boundary driven systems \cite{BermudezPlenio2013, RammHaffner2014, GuoPoletti2014}.    

Abelian gauge fields, like the magnetic field acting on electrons, are known to be able to steer the motion of particles while they are transported. The quantum Hall effect is a striking example of the effects of a gauge field on transport \cite{Klip, Laughlin}. In recent experiments with Josephson junctions it has been possible to modify heat transport with a magnetic field \cite{Giazotto2015}. Moreover, experimental advances have allowed the production of synthetic gauge fields with ultracold atoms \cite{DalibardOhberg2011, GoldmanSpielman2014}. A theoretical proposal has described how to produce synthetic gauge fields in microtrap arrays for ions \cite{BermudezPorras2011}. A simple geometry which can show the nontrivial effects of a gauge field is that of a ladder, i.e. two connected chains. This geometry has been extensively studied both in Josephson junctions arrays and bosonic atoms in optical lattices also in presence of interactions \cite{Kardar1986, Granato1990, DennistonTang1995, Nishiyama2000, OrignacGiamarchi2001, ChaShin2011, FazioVanDerZant2001, DharParamekanti2012, DharParamekanti2013, CrepinSimon2011, TovmasyanHuber2013, PetrescuLeHur2013, WeiMueller2014, AtalaBloch2014, TokunoGeorges2014, PiraudSchollwock2015, GreschnerVekua2015}. To be highlighted is the emergence of a quantum phase transition for bosonic particles, wich is characterized by the order parameter chiral current, defined as the difference of current between the upper and the lower chains. 
\begin{figure}
\includegraphics[width=\columnwidth]{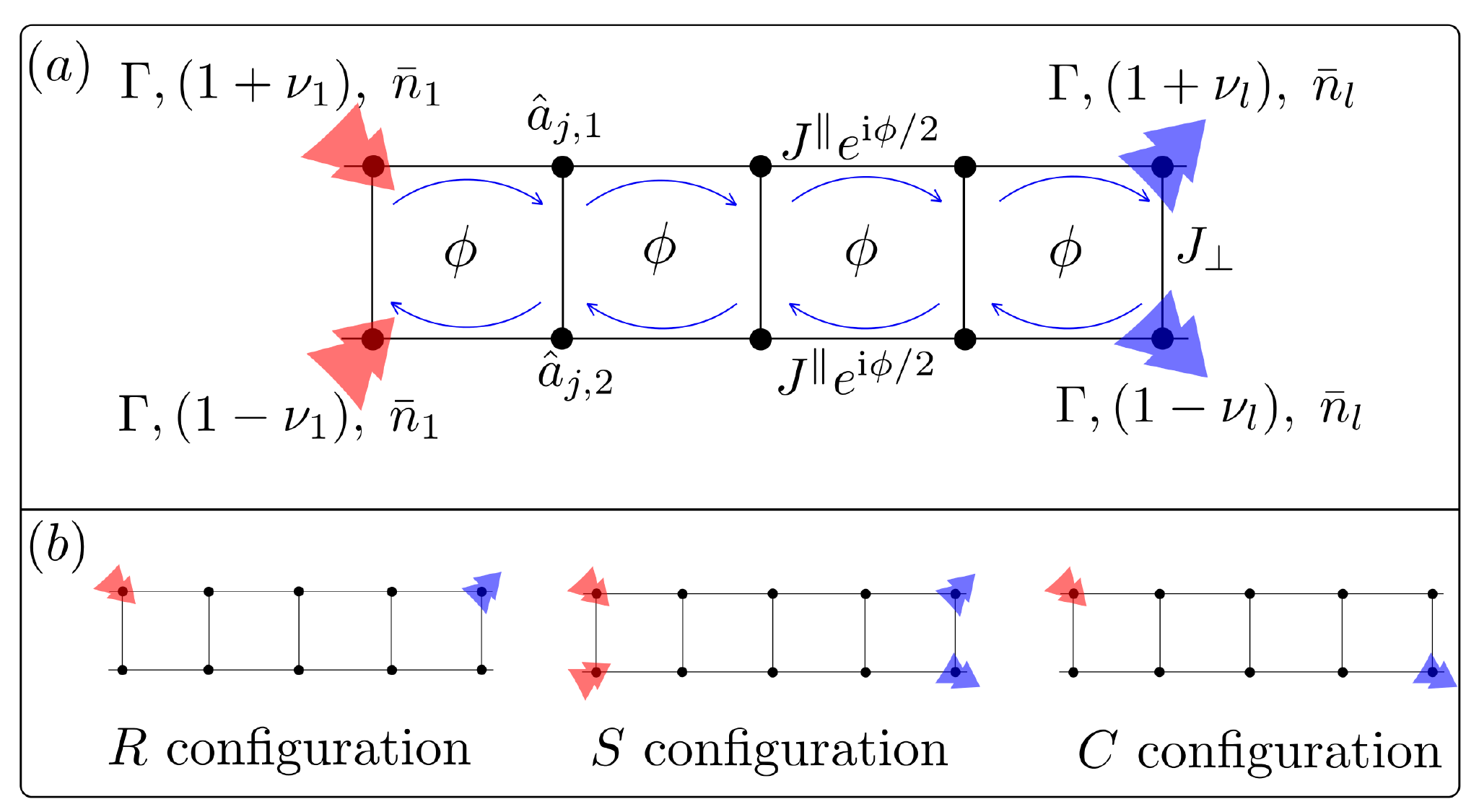}
\caption{(color online) (a) Ladder made of two coupled linear chains, referred to as legs of the ladder, with local bosonic excitations described by the annihilation operators at site $j$, $\aop_{j,p}$, where $p=1,2$ respectively for the upper and the lower leg. $\Jp$ is the tunnelling between the legs, on what are referred to as rungs of the ladder, while $\Jpa$ is the amplitude of tunnelling between sites in the legs. A gauge field imposes a phase $\phi$. The coupling to the baths is represented by the thick arrows. Each bath is characterized by the average density of bosons $\bar{n}_j$ the bath itself imposes on the rung $j$ and the strength of the coupling $\Gamma(1\pm\nu_j)$. 
(b) Three characteristic configurations: in the $R$ (reflection symmetric) configuration the baths are coupled only to the upper chain, in the $S$ (symmetric) configuration they are equally coupled to both chains while in the $C$ (centrosymmetric) configuration one bath is coupled to upper chain while the other to the lower one. Note that the configurations are only symmetric regarding the geometric coupling to the bath but not considering the baths' parameters. } \label{fig:model} 
\end{figure}
%

We study the interplay between the dissipation (and the geometry of its coupling to the system) which favours current flow and the gauge potential which tends to steer it. The gauge field can also cause significant changes in the energy levels' structure. We find that by tuning this field, combined with altering the geometry of the system-bath coupling, it is possible to drive non-equilibrium phase transitions which significantly affect the current flow both in its magnitude and in the pattern formed. This provides a mean to control the flow in this boundary driven systems. In particular, one phase transition occurs for the same parameters as the quantum phase transition although the emerging properties, characterized for instance by the chiral current, of the system are strikingly different in the dissipative or in the Hamiltonian systems. The other phase transition occurs when a gap opens in the spectrum of the Hamiltonian system. At this transition, depending on the geometry of the system-bath coupling, an abrupt change of the total current may occur. 
In section \ref{sec:model} we describe the model studied. We then analyze the total, section \ref{sec:total}, and chiral, section \ref{sec:chiral}, currents in the system. The current modulations are further discussed in section \ref{sec:momentum} and then we present symmetry considerations in section \ref{sec:symmetry}. In section \ref{sec:conclusions} we draw our conclusions.

\section{Model} \label{sec:model}
The system we study is represented in Fig.\ref{fig:model}: a ladder made of two coupled chains (or legs) of $l$ local bosonic modes which is coupled at its extremities to four baths. This could be realized with ion trap microarrays \cite{BermudezPorras2011} which are driven at the boundaries by side-band cooling \cite{LeibfriedWineland2003}. The evolution of an observable $\Op$ is given by 
\begin{equation}
\frac{d\Op}{dt} = - \frac{\im}{\hbar}\left[\Op,\Hop\right]+\Dop_H(\Op)\label{eq:MME}
\end{equation}   
where the Hamiltonian is given by 
\begin{align}
\Hop=-&\left(J^{\parallel} \sum_{p,j}\right.\allowbreak e^{\im(-1)^{p+1}\phi/2}\adop_{j,p}\aop_{j+1,p} \nonumber\\
&+ \allowbreak\left.\Jp\sum_{j} \adop_{j,1}\aop_{j,2} \right) + {\rm H.c.} 
\end{align}
Here $\Jpa$ is the tunnelling constant in the legs, $\Jp$ for the rungs, while $\aop_{j,p}$ ($\adop_{j,p}$) annihilates (creates) a boson in the upper (for $p=1$) or lower ($p=2$) chain at the $j$-th rung of the ladder. The tunnelling in the legs has a complex phase $\phi/2$ due to our choice of the gauge. 
The coupling to the baths is modelled by a dissipator 
\begin{align} 
\Dop_H(\Op)  = \sum_{j,p} \Gm_{(j,p),(j,p)} \allowbreak\left[ \allowbreak\bar{n}_{j,p} \left(\aop_{j,p}\Op\adop_{j,p} - \aop_{j,p}\adop_{j,p}\Op \right) \right. \nonumber \\
\allowbreak+  \left. (\bar{n}_{j,p} +1)\right. \allowbreak  \left.\left(\adop_{j,p}\Op\aop_{j,p} - \adop_{j,p}\aop_{j,p}\Op \right) \allowbreak + {\rm H.c.} \right]  
\end{align}
 in Lindblad form \cite{Lindblad1976, Gorini1976} where $j=1$ or $l$, $\Gm$ is a diagonal matrix, with diagonal elements $ \Gm_{(j,p),(j,p)}=\Gamma\left[1 - (-1)^{p} \nu_j\right]$, which means the coupling constant of the bosons at site $j$, and $\bar{n}_{j,p}$ is the local particle density that the dissipator would impose to the ion if isolated. $\Gamma$ is the overall coupling constant while $\nu_j$ is the asymmetric component of the dissipative coupling to the baths. Varying the values of $\nu_j$ alters the geometry of the system-bath coupling, making the coupling to the upper chain different from that to the lower one and gradually turns the system from the $S$ ($\nu_j=0$) to the $R$ or $C$ configurations depicted in Fig.\ref{fig:model}(b). As we shall see later in detail, changing $\nu_j$ (i.e. the geometry of the coupling) strongly affects the current flow generated in the ladder. 

\subsection{Solving the equation for the quadratic operator}
Now we introduce two diagonal matrix $\Lmp$ and $\Lmm$, which satisfy $\Lmp_{(j,p),(j,p)} = \Gm_{(j,p),(j,p)}\bar{n}_{j,p}$ and  $\Lmm_{(j,p),(j,p)} = \Gm_{(j,p),(j,p)}(\bar{n}_{j,p}+1)$ which implies that $\Gm = \Lmm - \Lmp$. We define another Hermitian matrix $\hm$ of which the non-zero elements are
\begin{align}
\hm_{(j,p),(j+1,p)} &= {\hm_{(j+1,p),(j,p)}}^{\ast} = J^{\parallel}e^{i(-1)^{p+1}\phi} \label{eq:h1}  \\
\hm_{(j,1),(j,2)} &= \hm_{(j,2),(j,1)} = J^{\perp} \label{eq:h2}
\end{align}
so that $\Hop$ can compactly be written as $\Hop = \sum_{(j,p),(k,q)}\hm_{(j,p),(k,q)}\aop_{j,p}^{\dagger}\aop_{k,q}$, where $1 \leq j, k \leq l$, and $p,q = 1, 2$. We focus on the quadratic observables $\Xm_{(j,p),(k,q)} = tr( \aop_{j,p}^{\dagger}\aop_{k,q} \rhop )$. Substituting $\Xm$ into the master equation Eq.(\ref{eq:MME}), we get
\begin{eqnarray}
 \frac{d \Xm}{dt} = -2\Pm \Xm - 2 \Xm \Pm^{\dagger} + 2\Lmp
\end{eqnarray}
where we have written $\Pm = (-\im\hm^t + \Gm )/2$. Therefore in the steady state we have
\begin{eqnarray}
\Pm \Xm + \Xm \Pm^{\dagger} = \Lmp.  \label{eq:pxxp}
\end{eqnarray}
We use this equation to compute the steady state properties of the system.

\begin{figure}
\includegraphics[width=\columnwidth]{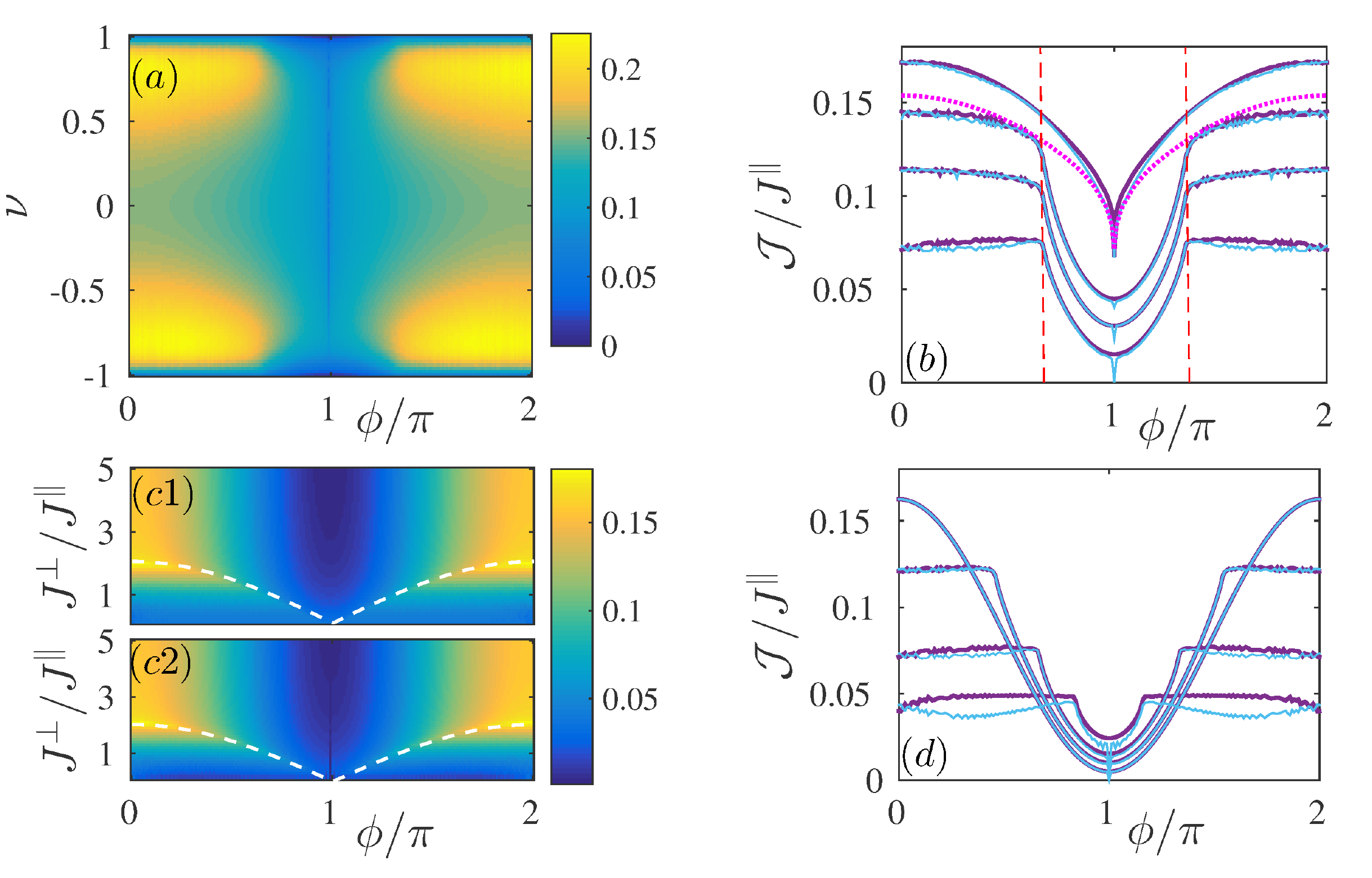}
\caption{(color online) (a) Total current in the ladder as a function of phase $\phi$ and asymmetric coupling $\nu$. In the positive range of $\nu$ we have chosen $\nu_1=\nu_l=\nu$ while in the negative range $\nu_1=-\nu_l=-\nu$. This implies that the line  $\nu_l=1$ corresponds to the $R$ configuration, $\nu_l=0$ at the $S$ configuration while $\nu_l=-1$ is the $C$ configuration. Here $\Jp=\Jpa$. In (b) the light blue curves are for $\nu_l>0$ while the dark purple for $\nu_l<0$. The values of $\nu_l$ are, from top to bottom, $\pm 0.4$, $\pm 0.96$, $\pm 0.98$ and $\pm 1$. The red vertical lines indicate the values of the phase $\phi=2\pi/3$ and $4\pi/3$ at which, as discussed later, a gap opens between the lower and upper band in the energy spectrum of the ladder. The light purple dotted line is for the $S$ configuration, i.e. $\nu=0$. (c) Depiction of the total current as a function of $\Jp$ and $\phi$ for the $R$ (c1) and $C$ (c2) configurations. The white dashed lines are given by Eq.(\ref{eq:totcurtransition}). (d) Selected curves from panels (c) showing $\cur$ versus $\phi$ for $\Jp/\Jpa=0.5,\;1,\;1.5,\;2.5$ (from bottom to top at $\phi=0$). The position of the transition shifts as $\Jp$ varies and it disappears for $\Jp\ge 2\Jpa$. The light blue and dark purple curves are for the $C$ and $R$ configurations respectively. In all figures shown the number of rungs $l=500$, $\bar{n}_1=0.5$, $\bar{n}_l=0.1$ and the strength of the coupling to the baths is $\Gamma=5\Jpa/\hbar$ (except if stated otherwise).} \label{fig:totcurrent} 
\end{figure}

\section{Total current}\label{sec:total} 
We study the current in the steady state of the system. The steady state is unique except for special cases (we will discuss this more in detail in section \ref{sec:symmetry}). The total current, $\cur=\sum_p\cur_{j,p}$, is given by the sum of the current in the legs $\cur_{j,p}$. The gauge invariant leg and rung particle currents are 
\begin{align}
\cur_{j,p} = \lan\im J^{\parallel}e^{\im(-1)^{p+1}\phi/2}\aop_{j,1}^{\dagger}\aop_{j+1,1} +{\rm H.c}\ran \label{eq:legcurrent}
\end{align}
and 
\begin{align} 
\cur_{j,1\rightarrow 2} = \lan \im J^{\perp} \aop_{j,1}^{\dagger}\aop_{j,2} + {\rm H.c.} \ran, \label{eq:rungcurrent}     
\end{align} 
derived from the continuity equations 
\begin{align}
\frac{\partial \lan\nop_{j,1}\ran}{\partial t} = \cur_{j-1,1} - \cur_{j,1} - \cur_{j,1\rightarrow 2} 
\end{align} and 
\begin{align}
\frac{\partial \lan\nop_{j,2}\ran}{\partial t} = \cur_{j-1,2} - \cur_{j,2} - \cur_{j,2\rightarrow 1} 
\end{align} for $1<j<l$.

Tuning the phase $\phi$ allows to vary the total current $\cur$ transported in the ladder. In particular, for sufficient coupling strength to the baths $\Gamma$, and {\it asymmetric configurations}, the current undergoes an abrupt change showing a non-equilibrium quantum phase transition. This is clearly depicted in Fig.\ref{fig:totcurrent}: in panels (a-b) we show the current $\cur$ versus the phase $\phi$ for different couplings to the baths, from the $S$ ($\nu=0$), to the $R$ ($\nu=1$) and $C$ ($\nu=-1$) configurations and intermediate cases. We stress two particular aspects: when the geometry of the coupling to the baths is close to the $R$ or $C$ configurations, the total current can be highly (or completely) suppressed around $\phi=\pi$. Moreover, there is a sharp change of the functional form of the current versus phase signalling the presence of a non-equilibrium phase transition. With Fig.~\ref{fig:totcurrent}(c,d) we explore the influence of the tunnelling between the legs $\Jp$. Fig.~\ref{fig:totcurrent}(c,d) show the shift, and eventually the dissapearence, of the transition as $\Jp$ is varied. The transition is demarked by the dashed white curve which is derived later to be Eq.(\ref{eq:totcurtransition}).   
\begin{figure}
\includegraphics[width=\columnwidth]{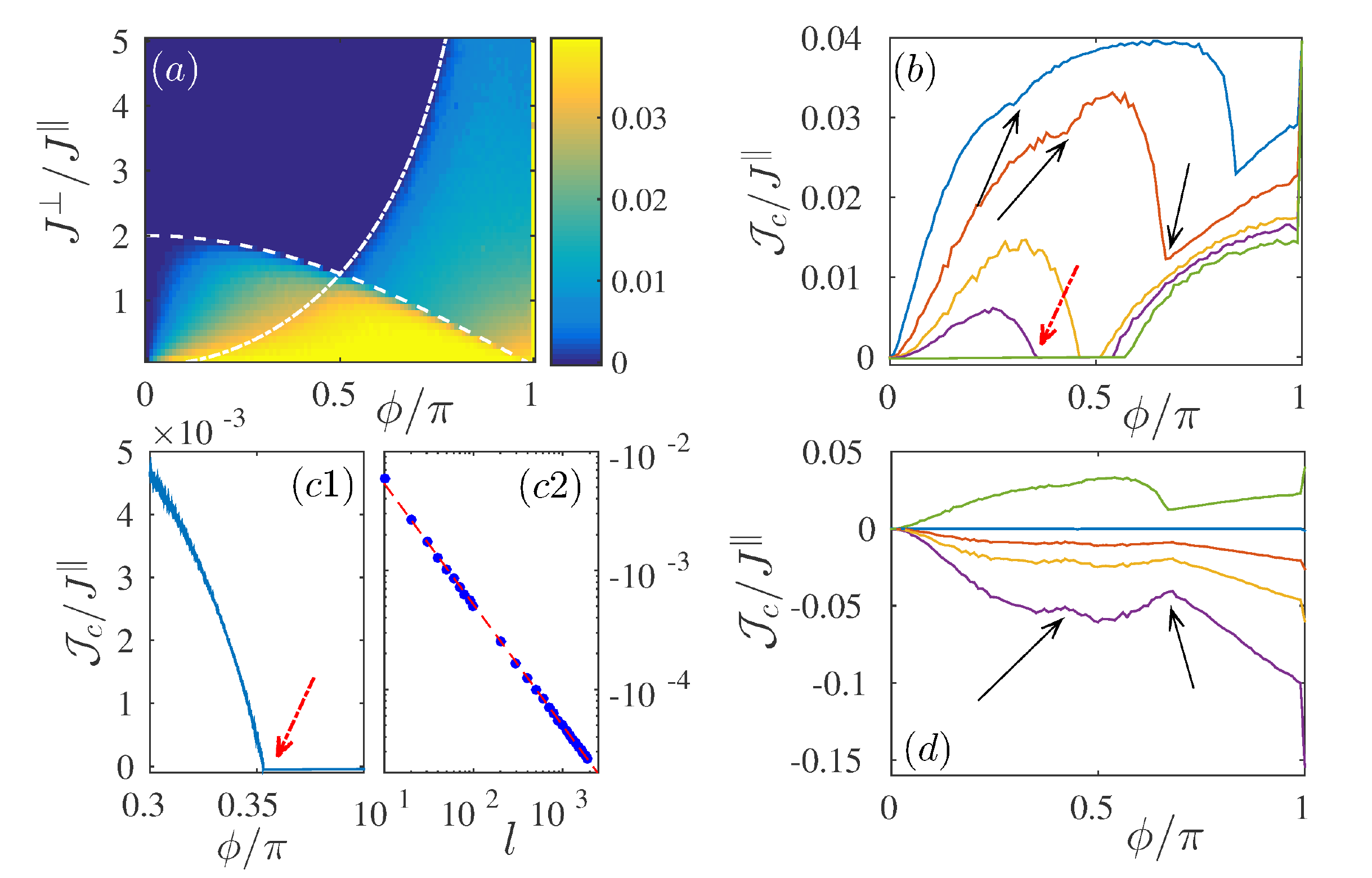}
\caption{(color online) (a) Chiral current as a function of perpendicular tunnelling $\Jp$ and phase $\phi$. (b) Chiral current versus $\phi$ for the selected vaules of $\Jp/\Jpa=0.5,\;1,\;1.5,\;1.7,\;2$ (from top to bottom). In (a) we show that the two transitions happen at values of the phase given by Eq.(\ref{eq:qptransition}), dash-dotted line, and (\ref{eq:totcurtransition}), dashed line. (c) Details of phase transition for $\Jp=1.7\Jpa$ in the $R$ configuration: (c1) A zoom-in of the transition for $l=1000$. (c2) Scaling of the chiral current versus the system size (number of rungs $l$) at the transition point $\phi\approx 0.3532...$, as given by Eq.(\ref{eq:totcurtransition}) [transition highlighted by the red-dashed arrow in Fig.\ref{fig:chiralcurrent}(b,c1)]. The circles are numerical data while the straight line is a power-law fit with exponent $-1$. (d) Chiral current versus phase $\phi$ for $\nu=\nu_1=\nu_l=0.02,\;0.18,\;0.38,\;0.78,\;1$  (from bottom to top) showing chiral current inversion. The arrows in (b) and (d) highlight some transitions.} \label{fig:chiralcurrent} 
\end{figure}
\noindent

\section{Chiral current}\label{sec:chiral} 
An analysis of the chiral current 
\begin{align}
\cur_c=\sum_j(\cur_{j,1}-\cur_{j,2})/l \label{eq:chiralcurrent}  
\end{align}
shows that, unlike the unitary case without baths for which only one transition occurs, there can be two transitions. Moreover, symmetry guarantees that the chiral current vanishes for the symmetric coupling to the baths $S$ and the centrosymmetric coupling $C$ and scenarios in between, as shown in Sec. \ref{sec:symmetry}. Results for the $R$ configuration are depicted in Fig.\ref{fig:chiralcurrent}(a,b) where we show the combined effect of $\Jp$ and $\phi$. At a smaller value of $\Jp$, the chiral current presents two abrupt changes in its derivative with respect to $\phi$, [some of these points are highlighted with arrows in Fig.\ref{fig:chiralcurrent}(b)], then, at intermediate values of $\Jp$ two regions with chiral current are clearly separated and drift apart until, at large values of $\Jp$, the chiral current is only present near $\phi=\pi$. A detail of one transition is shown in Fig.\ref{fig:chiralcurrent}(c1). In Fig.\ref{fig:chiralcurrent}(c2) we show that the chiral current goes to $0$ as a power law at the transition point as the system's size increases, signalling a phase transition. In Fig.\ref{fig:chiralcurrent}(d) we study the role of the amplitude of the asymmetric coupling $\nu$ as the system varies from the $S$ towards the $R$ configurations while keeping $\Jp=\Jpa$. We also note a chiral current inversion for large enough $\nu$, which is clearly shown in Fig.\ref{fig:chiralcurrent}(d). Unlike in \cite{GreschnerVekua2015}, the chiral current reversal is due to the interplay between the dissipative and the Hamiltonian dynamics.           
\begin{figure}[t]
\includegraphics[width=\columnwidth]{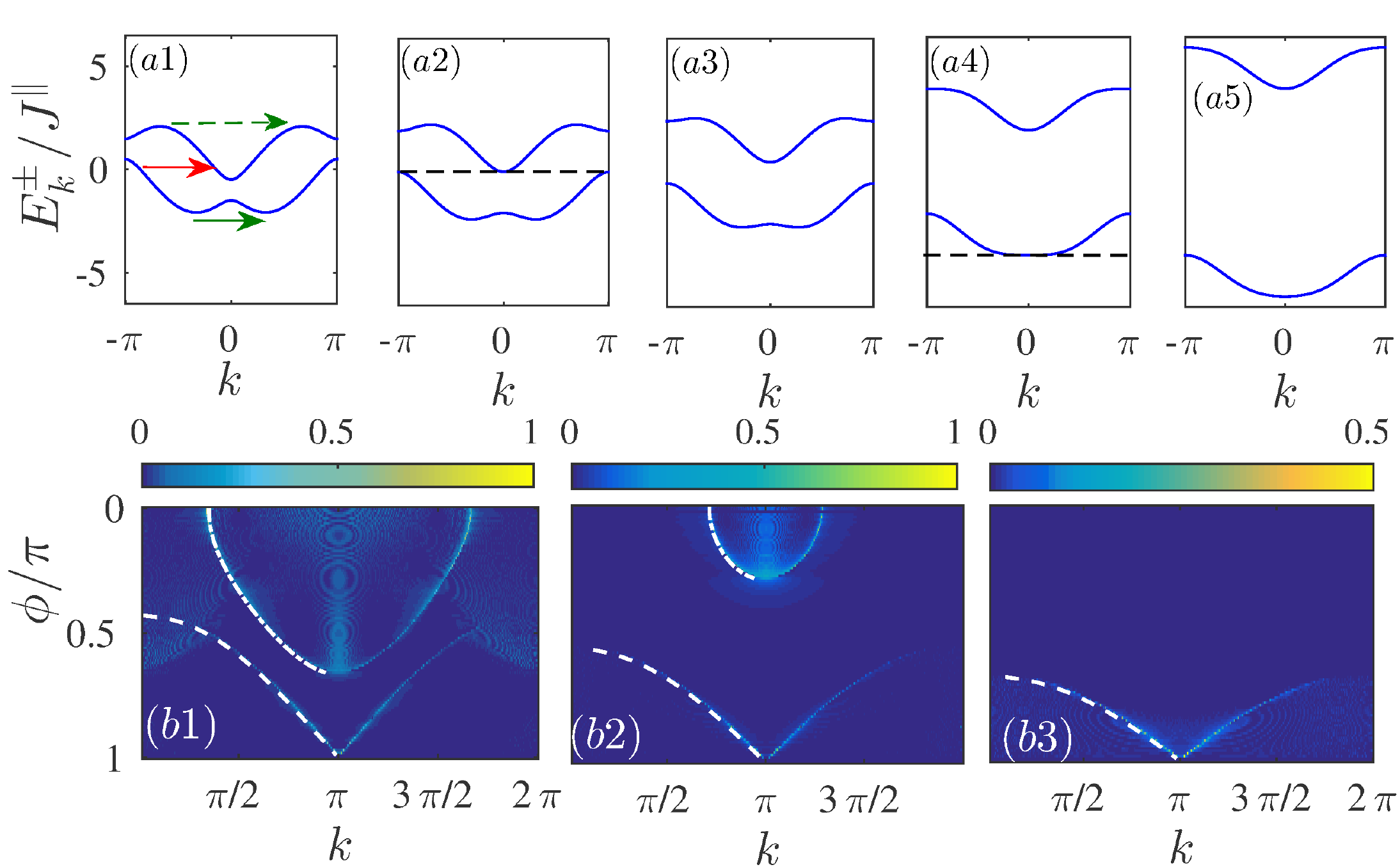}
\caption{(color online) (a1-5) Energy spectrum of the ladder for phase $\phi=2\pi/3$ and the amplitude of tunnelling in the bulk respectively $\Jp/\Jpa=0.5,\;1,\;1.5,\;3,\;5$. The dashed line in (a2) highlights that this corresponds to the critical value of $\Jp$ at which the gap between the lower and upper band is $0$. The dashed line in (a4) points out that $\Jp=3\Jpa$ is the critical value at which the minima have merged. (b1-b3) Momentum probability distribution for the current in the lower leg as a function of $\phi$ and the momentum $k$ for the $R$ configuration (other configurations, except the symmetric $S$ configuration present a similar structure). In the three plots $\Jp/\Jpa$ is respectively $1$, $1.8$ and $3$. The white dashed lines are analytically obtained from Eq.(\ref{eq:k_intra}) and highlight current due to bath induced intra-band coupling as, for example, depicted in (a1) by the green arrows. The white continuous lines are given by Eq.(\ref{eq:k_inter}) and are due to inter-bands coupling [one example is the red arrow in (a1)]. Note that the point $k=0$ has been removed to make the patterns more visible.} \label{fig:spectra} 
\end{figure}

\section{Momentum distribution}\label{sec:momentum}   
To gain deeper insight into the distribution of the current flow in the system we study the Fourier transform of the current in the lower leg $\cur^{k,2}$. In Fig.\ref{fig:spectra}(b1-b3) we depict the typical results for $|\cur^{k,2}|$ (excluding $k=0$). If $|\cur^{k,2}|$ is zero for any $k$ it means that the current is uniform within the leg which implies no rung current. Instead the presence of peaks results in modulating patterns in the rung current. To gain an understanding of the spatial modulation we consider the spectrum for an infinite ladder with periodic boundary conditions. The quadratic Hamiltonian can be readily diagonalized as \begin{align}
\Hop=\sum_k E^{\pm}_k \alpop^{\dagger}_{k,\pm} \alpop_{k,\pm}
\end{align} 
with 
\begin{align}
E^{\pm}_k =& -2\Jpa\cos(\phi/2)\cos(k) \nonumber \\
&\pm \sqrt{J^{\perp\;2} + \left[2\Jpa\sin(\phi/2)\sin(k)\right]^2}
\end{align}    
and $\alpop_{k,\pm}$ the annihilation operator of a particle at momentum $k$ in the upper ($+$) or lower ($-$) band. 
The spectrum  presents three typical scenarios, Figs.\ref{fig:spectra}(a1, a3, a5), separated by the transition points in Figs.\ref{fig:spectra}(a2, a4). In Fig.\ref{fig:spectra}(a1) the two bands overlap and each band has either two minima (lower band) or two maxima (higher band); In Fig.\ref{fig:spectra}(a3) the two bands do not overlap while each band still has either two minima or two maxima; In Fig.\ref{fig:spectra}(a5) each band has only one minimum (lower band) or maximum (higher band). In Fig.\ref{fig:spectra}(a2) the two bands separate while in Fig.\ref{fig:spectra}(a4) there is a transition from two minima (maxima) to one minimum (maximum).

We can now intuitively understand the occurrence and spatial period of the modulation by a perturbative analysis of the system for small coupling to the bath $\Gamma$ (in a parallel way to \cite{ZnidaricProsen2011}. At zero-th order, any single particle correlator of the type $\lan\alpop^\dagger_{k,\sigma} \alpop_{k',\sigma'}\ran$ is constant in time as long as $E^{\sigma}_{k}=E^{\sigma'}_{k'}$. The dissipation, being local, can possibly couple all different momenta, and, in first order, it acts strongly on this constant manifold. It thus couple different energetically degenerate momenta, whether in the same band (intra-band) or in different bands (inter-band). This results in a modulation given by the difference or sum of the two momenta $k$ and $k'$. Hence the maxima of $|\cur^{k,2}|$ occur at    
\begin{align}
k_{{\rm intra}}&= \pm 2{\rm arccos}\left(\theta_{\rm intra}\right) \label{eq:k_intra} \\
k_{{\rm inter}}&=\pm{\rm arccos}\left(\theta_{\rm inter}\right) \label{eq:k_inter}     
\end{align} 
respectively from the intra-band and inter-band coupling. Here 
\begin{align}
\theta_{\rm intra}=\cot\left(\phi/2\right)\sqrt{\sin^2\left(\phi/2\right)+ \left[\Jp/(2\Jpa)\right]^2}
\end{align} 
 and 
\begin{align} 
 \theta_{\rm inter}=2\cos\left(\phi/2\right)\left[\cos\left(\phi/2\right)-\left(\Jp/(2\Jpa)\right)\right]    -1. 
 \end{align}       
It is at the occurrence of degeneracies that a phase transition manifests itself, provided that the coupling via the bath is non-zero. The critical values of $\Jp/\Jpa$ and $\phi$ for the transition lines can be readily computed from the analytical expression of the spectrum, resulting in 
\begin{align}
\bar{J}_{\perp}&=2\Jpa\cos\left(\bar{\phi}/2\right), \label{eq:totcurtransition} \\  
\tilde{J}_{\perp}&=2\Jpa\tan\left(\tilde{\phi}/2\right)\sin\left(\tilde{\phi}/2\right) \label{eq:qptransition}               
\end{align}
respectively for the gap opening (the condition being $E^+_0=E^-_{\pi}$) and for the transition between one to two minima. 
The phase transitions lines are highlighted in Fig.\ref{fig:chiralcurrent}(a) by the white dashed curve, Eq.(\ref{eq:totcurtransition}), and the dot-dashed white curve, Eq.(\ref{eq:qptransition}). Moreover the opening of the gap also results in an abrupt reduction of the total current as shown by the dashed curves in Fig.\ref{fig:totcurrent}(c1-c2). 
It is easy to show that in the symmetric, $S$, configuration, the coupling via the bath of the $0$ and $\pi$ momenta vanishes which results in a smooth change of the total current and no phase transition occurs (see the light purple dotted line in Fig.\ref{fig:totcurrent}(b)).  


\section{Symmetry considerations}\label{sec:symmetry} 
Here we discuss some properties of the system which can be derived from symmetry considerations. We will first show the effect of changing $\phi\rightarrow -\phi$, section \ref{sec:mphi}, then of linear changes of the baths parameters $\bar{n}_{i,p}$, section \ref{sec:shift}, symmetry properties of the total and chiral current depending on the configuration, \ref{sec:configs}, and last for the case of $\phi=\pi$, sections \ref{sec:evenrungs} and \ref{sec:oddrungs}. 


\subsection{Change of phase from $\phi \rightarrow -\phi$} \label{sec:mphi}
Let us consider a transformation from $\phi \rightarrow -\phi$. The dissipation is untouched, while the new Hamiltonian $\hat{H}^{\prime}$ is such that $\hat{H}^{\prime}=\Hop^*$ (the complex conjugate of $\Hop$). We thus have that $\hm^{\prime} = \hm^{\ast}$. Denoting the quadratic observables of the new system as $\Xm^{\prime}$, using the particular form of $\hm$ and the fact that $\Lmp$ and $\Lmm$ are diagonal, it can be proved (see Appendix) that 
\begin{eqnarray}
\Xm^{\prime}_{(j,p),(j+k,p+q)} = (-1)^{k+q}\Xm_{(j,p),(j+k,p+q)}^{\ast} \label{eq:xprimeximportant}
\end{eqnarray}
This implies that, unlike in the unitary system for which the current is antisymmetric for a change of $\phi\rightarrow -\phi$, in the dissipative case it is symmetric. 
For instance the rung current for a phase $-\phi$, from Eq.(\ref{eq:rungcurrent}), 
\begin{align}
\cur_{j,1\rightarrow 2}(-\phi)&=-2\Jp\mathcal{I} \left(\Xm'_{(j,1),(j,2)}\right)\nonumber\\ &=-2\Jp\mathcal{I} \left(\Xm_{(j,1),(j,2)}\right)\nonumber \\ 
&=\cur_{j,1\rightarrow 2}(\phi) \label{eq:rungcurphi}        
\end{align}         
where $\mathcal{I}(.)$ takes the imaginary part of its argument. Similarly for the leg current, from Eq.(\ref{eq:legcurrent}),   
\begin{align}
\cur_{j,p}(-\phi)&=-2J^{\parallel}\mathcal{I} \left(e^{-\im(-1)^{p+1}\phi/2}\Xm'_{(j,p),(j+1,p)}\right)\nonumber\\ &=-2J^{\parallel}\mathcal{I} \left(e^{\im(-1)^{p+1}\phi/2}\Xm_{(j,p),(j+1,p)}\right)\nonumber\\ 
&=\cur_{j,p}(\phi)  \label{eq:legcurphi}   
\end{align}

\subsection{Linear change of the baths parameters $\bar{n}_{j,p}$} \label{sec:shift}    
Let us consider a change in $\Xm$ such that $\Xm^{\prime} = \Xm +s \mathbb{1}$, where $\mathbb{1}$ means the identity matrix. It is straightforward, from Eq.(\ref{eq:pxxp}), to see that $\Xm^{\prime}$ satisfies the equation 
\begin{eqnarray}
\Pm\Xm' + \Xm \Pm^{\dagger} = {\Lmp}^{\prime} \label{eq:shiftX}  
\end{eqnarray} 
with ${\Lmp}^{\prime}=\Lmp+s\Gm$. 
The consequences of Eq.(\ref{eq:shiftX}) is apparent. $\Xm^{\prime}$ will be a solution to another system with $\Gm^{\prime} =\Gm$ but with an average occupation shifted as $\bar{n}^{\prime}_{j,p} = \bar{n}_{j,p} + s$. A change in the average occupation fixed by the dissipation thus only change the local occupations but not the currents nor the kinetic energy. 

In a similar manner it is also easy to realize that $\Xm^{\prime} = s \Xm$ is the solution to the system with the same $\hm$ and $\Gm$, but $\bar{n}^{\prime}_{j,p} = s\bar{n}_{j,p}$.
A natural consequence of the results in this subsection is that it is possible to compute the observables for the case in which one boundary is set to $\bar{n}_{j,p}=0$ while the other to any value and then compute all the other possible cases by a shift or a dilation in $\bar{n}_{j,p}$.

\subsection{Symmetries of total and chiral currents in the various configurations}\label{sec:configs}
In this section we analyse the symmetry of both the Hamiltonian and the dissipation. First, it is easy to see that if there is a unitary transformation $\Um$ such that $\Xm^{\prime} = \Um^{\dagger} \Xm \Um$, which is $\Xm = \Um \Xm^{\prime} \Um^{\dagger}$, then $\Xm^{\prime}$ satisfies
\begin{eqnarray}
\Um^{\dagger}\Pm \Um \Xm^{\prime} + \Xm^{\prime} \Um^{\dagger} \Pm^{\dagger} \Um = \Um^{\dagger} \Lmp \Um,
\end{eqnarray}
which means that $\Xm^{\prime}$ is the solution to the system with $\Pm^{\prime} = \Um^{\dagger}\Pm \Um$, ${\Lmp}^{\prime} = \Um^{\dagger}\Lmp \Um$.

\subsubsection{$R$ configuration}    
In the $R$ configuration, we define the transformation $\Um$ such that it has non-zero elements only on the anti-diagonal, $\Um_{(i,p),(l-i+1,p)} = 1$, then we have
\begin{eqnarray}
\hm^{\prime} = \Um^{\dagger} \hm \Um = \hm^{\ast};\\
\Gm^{\prime} = \Um^{\dagger} \Gm^{\prime} \Um = \Gm
\end{eqnarray} 
and ${\Lmp}^{\prime} = \Um^{\dagger} \Lmp \Um$ satisfies ${\Lmp}^{\prime}_{(1,1),(1,1)} = \Lmp_{(l,1),(l,1)}$ and ${\Lmp}^{\prime}_{(l,1),(l,1)} = \Lmp_{(1,1),(1,1)}$. From section \ref{sec:shift}, we know that the off diagonal terms of $\Xm$ are invariant to a shift of $\Lmp$ by a scalar. So we can just set $\Lmp_{(1,1),(1,1)} = 0$, and this will not affect the off diagonal terms. Then we find that ${\Lmp}^{\prime}$ and $\Lmp$ are related by
\begin{eqnarray}
{\Lmp}^{\prime} = -\Lmp + \Lmp_{(l,1),(l,1)} \one,
\end{eqnarray}
Denoting $\Pm_{\phi}^{\prime}=\Pm_{-\phi}$ (as $\Pm$ is a function of $\Hop$, it is also a function of $\phi$ which we now show explicitly), we can thus write 
\begin{align}
\Pm_{\phi}^{\prime}{\Xm}^{\prime} + {\Xm}^{\prime} {\Pm_{\phi}^{\dagger}}^{\prime} &= {\Lmp}^{\prime} \nonumber\\ 
\Pm_{-\phi}{\Xm}^{\prime} + {\Xm}^{\prime} {\Pm_{-\phi}^{\dagger}} &= -{\Lmp} + \Lmp_{(l,1),(l,1)} \one. \label{eq:RconfPprime}   
\end{align}    
Eq.(\ref{eq:RconfPprime}), together with Eqs.(\ref{eq:rungcurphi},\ref{eq:legcurphi}) and considering that the current is independent of a constant shift in $\Lmp$, results in    
\begin{eqnarray}
\cur_{l+1-i,1\rightarrow 2} = -\cur_{i,1\rightarrow 2}
\end{eqnarray}
and 
\begin{eqnarray}
\cur_{l-i,p} = \cur_{i,p}. 
\end{eqnarray}

\subsubsection{$C$ configuration}
For the $C$ configuration the system Hamiltonian is invariant under the transformation $i \rightarrow l+1-i$, $p\rightarrow 3-p$, and the dissipation will be reverted. 
Therefore $\Pm_{\phi}^{\prime}=\Pm_{\phi}$ and we can write 
\begin{align}
\Pm_{\phi}^{\prime}{\Xm}^{\prime} + {\Xm}^{\prime} {\Pm_{\phi}^{\dagger}}^{\prime} &= {\Lmp}^{\prime} \nonumber\\ 
\Pm_{\phi}{\Xm}^{\prime} + {\Xm}^{\prime} {\Pm_{\phi}^{\dagger}} &= -{\Lmp} + \Lmp_{(l,1),(l,1)} \one. \label{eq:CconfPprime}   
\end{align}    
Again, since the current is independent of a constant shift of $\Lmp$, we can state that 
\begin{eqnarray}
\cur_{l+1-i,1\rightarrow 2} = \cur_{i,1\rightarrow 2} 
\end{eqnarray}
and 
\begin{eqnarray}
\cur_{l-i,3-p} = \cur_{i,p}
\end{eqnarray}
We can hence derive that the chiral current $\cur_c=0$ in the $C$ configuration.

\subsubsection{$S$ configuration}
In this case the system has the symmetry of both the $R$ configuration and the $C$ configuration. This means we have
\begin{eqnarray}
\cur_{l-i,3-p} = \cur_{i,p} = \cur_{i,3-p},
\end{eqnarray}
which also implies that the current on the rungs $\cur_{i,1\rightarrow 2}$, and the chiral current are 0.

\subsection{$\phi=\pi$ and even number of rungs} \label{sec:evenrungs}
For an even number of rungs, $l = 2N$, we can arrange the Hamiltonian such that
\begin{align}
\Hop &= \{\aop_{1,1},\aop_{3,1},\dots, \aop_{2N-1,1},\aop_{2,2},\aop_{4,2},\dots ,\aop_{2N,2} \}\times \nonumber \\
&\Hm \{\adop_{1,2},\adop_{3,2},\dots , \adop_{2N-1,2}, \adop_{2,1},\adop_{4,1},\dots , \adop_{2N,1} \} + {\rm H.c.} \label{eq:grouping} 
\end{align}
The coefficient matrix $\Hm$ can be written as
\begin{eqnarray}
\Hm	 = \left(
 	\begin{array}{cc}
 	\Am  & \Bm \\
    \Cm & \Dm \\
    \end{array}
	\right), \label{eq:Hm} 
\end{eqnarray}
where $\Am$,$\Bm$,$\Cm$,$\Dm$ are $N \times N $ matrix with the only non-zero components being $\Am_{j,j} = J^{\perp}$, $\Dm_{j,j} = J^{\perp}$, $\Bm_{j,j} = J^{\parallel} e^{-\im\phi/2}$, $\Bm_{j+1,j} = J^{\parallel}e^{\im\phi/2}$, $\Cm_{j,j} = J^{\parallel}e^{-\im\phi/2}$, $\Cm_{j,j+1} = J^{\parallel} e^{\im\phi/2}$. Considering the singular value decomposition $\Hm = \Um\Sm\Vm^{\dagger}$, then we can perform two unitary transformation to the system as
\begin{align}
\{\aop_{1,1},\aop_{3,1},\dots, \aop_{2N-1,1},\aop_{2,2},\aop_{4,2},\dots ,\aop_{2N,2} \} = \bop_{1 \rightarrow 2N} \Um^{\dagger} \nonumber \\
\{\adop_{1,2},\adop_{3,2},\dots , \adop_{2N-1,2}, \adop_{2,1}, \adop_{4,1},\dots , \adop_{2N,1} \} = \Vm  \cdop_{1 \rightarrow 2N} \nonumber
\end{align} 
Writing $\Hop$ in terms of these $4N$ new modes $\bop_{1 \rightarrow 2N}$, $\cop_{1 \rightarrow 2N}$, we get
\begin{eqnarray}
\Hop = \sum_{j=1}^{2N}\Sm_{j,j}\bop_j \cdop_j + {\rm H.c.} 
\label{eq:svd}
\end{eqnarray}
When $\phi=\pi$, the tunneling terms $J^{\parallel}_{i,1} = -J^{\parallel}_{i,2}$ which implies that $\Bm = -\Cm^{\dagger}$. This results in $\Hm^{\dagger}\Hm$ and $\Hm\Hm^{\dagger}$ to be block diagonal. Hence we can choose the left transformation $\Um$ to be block diagonal 
which means that 
$\{\aop_{1,1},\aop_{3,1},\dots, \aop_{2N-1,1}\}$ will belong to one block while $\{\aop_{2,2},\aop_{4,2},\dots ,\aop_{2N,2} \}$ to another. In the $C$ configuration, the jump operator $\aop_{1,1}$ and $\aop_{l,2}$ each belong to a different block. 
Hence there will be no connections between the two sub blocks, and each sub system will reach steady state independentely. 
This exact symmetry for $\phi=\pi$ and even number of rungs has an interesting consequence. In fact this implies that, even as $l$ increases the value of the total current for even or odd number of sites ladders will have a finite difference.

\subsection{$\phi=\pi$ and odd number of rungs} \label{sec:oddrungs}     
In case $l=2N+1$ we can arrange the Hamiltonian similarly to Eq.(\ref{eq:grouping})  
where the coefficient matrix $\Hm$ has an identical structure to Eq.(\ref{eq:Hm}).   
However, in this case $\Am$ is an $(N+1)\times (N+1)$ matrix, $\Bm$ is $(N+1) \times N$, $\Cm$ is $N \times (N+1)$, and $\Dm$ is $N \times N$, whose elements are the same as for the even case. Hence, also in this case, when $\phi=\pi$, $\Bm = -\Cm^{\dagger}$, which implies that the Hamiltonian can be written as two blocks. For the $R$ configuration only one block is connected to the the baths, which 
%
%
%
implies the existence of $2N$ modes dark modes. 


\section{Conclusions}\label{sec:conclusions} 
We have analyzed a rich quantum system in which, depending of the Hamiltonian parameter, there can be a quantum phase transtion and a gap opening. We have shown that these two features can be used to gain a high degree of control of the transport properties of the system when it is coupled to baths at its boundaries. Most importantly, we have shown how the characteristics of the system can change significantly depending on the geometry of the system-bath coupling. 
In particular, we have shown that the interplay between (1) the current-imposing baths at the boundaries, (2) their geometrical coupling to the ladder and (3) a gauge field which tends to alter the current flow and causes significant changes in the energy spectrum of the bulk system, results in non-equilibrium phase transitions in which the magnitude and the pattern of the current transported can change significantly. This phenomenology extends and differs significantly from the one studied in the groundstate of the Hamiltonian system.  
We have shown the presence of two regions in which chiral currents can emerge and we have connected the boundaries of these regions with (i) the emergence of two minima in the energy spectrum and (ii) an opening of a gap between the lower and higher bands. The opening of the gap also causes a significant reduction of the total current transported for asymmetric system-bath couplings.  
We have also complemented our analysis with symmetry considerations which allow us to show that chiral current is non-zero only in the $R$ configuration, and the peculiar behavior for $\phi=\pi$. 
These results shed light on general ways to manipulate and control transport in quantum systems thanks to quantum phase transitions and opening of gaps in the bulk spectrum of the system. In future works, the role of the geometry of the system-bath coupling could be studied further as a mean to generate useful non-equilibrium quantum phases.

\section*{Akwnoldgements} 
We acknowledge insightful discussions with U. Bissbort, R. Fazio, T. Giamarchi, B. Gr\'emaud, C. Kollath, M. Mukherjee and S. Yang. D.P. acknowledges support from AcRF MOE Tier-II (project MOE2014-T2-2-119, WBS R-144-000-350-112) and AcRF MOE Tier-I (project SUTDT12015005).

\begin{appendix}

\section{Details of inversion $\phi\rightarrow -\phi$}
In this appendix we will prove Eq.(\ref{eq:xprimeximportant}) in the main text.
First it is useful to write Eq.(\ref{eq:pxxp}) in a different form 
\begin{equation}
\hm^t \Xm - \Xm \hm^t = \im \left(2 \Lmp - \Gm \Xm - \Xm \Gm \right)\label{eq:hxxh}
\end{equation}
We consider another system whose dissipation is the same while the Hamiltonian $\hm^{\prime} = \hm^{\ast}$ (due, for instance, to an inversion of $\phi\rightarrow -\phi$). The steady state of this system, which we assume to be unique as for the original system $\hm$, will be $\Xm^{\prime}$.  $\Xm^{\prime}$ will then be the solution to the equation
\begin{equation}
\hm^{\dagger} \Xm^{\prime} - \Xm^{\prime} \hm^{\dagger} = \im \left(2 \Lmp - \Gm \Xm^{\prime} - \Xm^{\prime} \Gm \right)
\end{equation}
Taking complex conjugation of both side, and writing $\Xm^{\prime\prime} = {\Xm^{\prime}}^{\ast}$, we get
\begin{eqnarray}
\hm^t \Xm^{\prime\prime} - \Xm^{\prime\prime} \hm^t =  -\im \left(2 \Lmp - \Gm \Xm^{\prime\prime} - \Xm^{\prime\prime} \Gm \right)\label{eq:hxprimeprimeh}
\end{eqnarray}
We now show that 
\begin{eqnarray}
\Xm^{\prime\prime}_{(j,p),(j+k,p+q)} &=& (-1)^{k+q} \Xm_{(j,p),(j+k,p+q)} \label{eq:Xtransform}
\end{eqnarray}
where $\Xm$ solves Eq.(\ref{eq:pxxp}) and hence Eq.(\ref{eq:hxxh}). 
To prove the validity of this ansatz, we substitute Eq.(\ref{eq:Xtransform}) in the left-hand side of Eq.(\ref{eq:hxprimeprimeh}) and get  
\begin{align}
&(h^t \Xm^{\prime\prime} - \Xm^{\prime\prime}h^t)_{(i,m),(j,n)} \nonumber \\ 
&= \sum_{(k,p)}\hm_{(k,p),(i,m)}\Xm^{\prime\prime}_{(k,p),(j,n)} - \Xm^{\prime\prime}_{(i,m),(k,p)}\hm_{(j,n),(k,p)} \nonumber \\
&= \hm_{(i-1,m),(i,m)}\Xm^{\prime\prime}_{(i-1,m),(j,n)} + \hm_{(i+1,m),(i,m)}\Xm^{\prime\prime}_{(i+1,m),(j,n)} \nonumber \\
 &+ \hm_{(i,3-m),(i,m)}\Xm^{\prime\prime}_{(i,3-m),(j,n)} - \Xm^{\prime\prime}_{(i,m),(j-1,n)}\hm_{(j,n),(j-1,n)} \nonumber \\
 &- \Xm^{\prime\prime}_{(i,m),(j+1,n)}\hm_{(j,n),(j+1,n)} - \Xm^{\prime\prime}_{(i,m),(j,3-n)}\hm_{(j,n),(j,3-n)} \nonumber \\ 
 &= (-1)^{j-i+1+n-m}\hm_{(i-1,m),(i,m)}\Xm_{(i-1,m),(j,n)} \nonumber \\
 &+ (-1)^{j-i-1+n-m}\hm_{(i+1,m),(i,m)}\Xm_{(i+1,m),(j,n)} \nonumber \\ 
 &+ (-1)^{j-i+n+m-3}\hm_{(i,3-m),(i,m)}\Xm_{(i,3-m),(j,n)} \nonumber \\
 &- (-1)^{j-i-1+n-m}\Xm_{(i,m),(j-1,n)}\hm_{(j,n),(j-1,n)} \nonumber \\  
 &- (-1)^{j-i+1+n-m}\Xm_{(i,m),(j+1,n)}\hm_{(j,n),(j+1,n)} \nonumber \\ 
 &- (-1)^{j-i+3-n-m}\Xm_{(i,m),(j,3-n)}\hm_{(j,n),(j,3-n)} \nonumber \\ 
 &= (-1)^{j-i+n-m-1}\left[\hm_{(i-1,m),(i,m)}\Xm_{(i-1,m),(j,n)}\right. \nonumber \\
 &+ \hm_{(i+1,m),(i,m)}\Xm_{(i+1,m),(j,n)} + \hm_{(i,3-m),(i,m)}\Xm_{(i,3-m),(j,n)} \nonumber \\
 &- \Xm_{(i,m),(j-1,n)}\hm_{(j,n),(j-1,n)} - \Xm_{(i,m),(j+1,n)}\hm_{(j,n),(j+1,n)} \nonumber \\
 &- \left.\Xm_{(i,m),(j,3-n)}\hm_{(j,n),(j,3-n)}\right] \nonumber \\ 
 &= (-1)^{j-i+n-m-1} (\hm^t \Xm - \Xm \hm^t)_{(i,m),(j,n)}\label{eq:xprimeleft}
\end{align}
For a portion of the right-hand side of Eq.(\ref{eq:hxprimeprimeh}) we get
\begin{align}
&(-\Gm \Xm^{\prime\prime}-\Xm^{\prime\prime}\Gm)_{(i,m),(j,n)} \nonumber \\ 
&= (-1)^{j-i+n-m}(-\Gm_{(i,m),(i,m)}- \Gm_{(j,n),(j,n)})X_{(i,m),(j,n)} \nonumber
\end{align} 
Therefore, using the fact that $\Lmp$ is diagonal, we have that the full right-hand side of Eq.(\ref{eq:hxprimeprimeh}) becomes 
\begin{align}
& -\im(2\Lmp - \Gm \Xm^{\prime\prime}-\Xm^{\prime\prime}\Gm)_{(i,m),(j,n)} \nonumber \\
&=  \im(-1)^{j-i+n-m-1}(2\Lmp -\Gm \Xm-\Xm\Gm)_{(i,m),(j,n)}\label{eq:xprimeright} 
\end{align}
Combining Eqs.(\ref{eq:hxxh}, \ref{eq:xprimeleft}, \ref{eq:xprimeright}), we can see that Eq.(\ref{eq:Xtransform}) is indeed a solution of Eq.(\ref{eq:hxprimeprimeh}). Since we have assumed that the system has unique steady state, and $\Xm^{\prime\prime}$ is the complex conjugate of $\Xm^{\prime}$, we have proved Eq.(\ref{eq:xprimeximportant}) in the main text.


In case that the Hamiltonian is real, we know that $\Xm^{\prime} = \Xm$, which means
\begin{eqnarray}
\Xm_{(j,p),(j+k,p+q)} = (-1)^{k+q}\Xm_{(j,p),(j+k,p+q)}^{\ast}.
\end{eqnarray} 
Therefore we have $\Xm_{(j,p),(j+k,p)} = (-1)^{k}\Xm_{(j,p),(j+k,p)}^{\ast}$, which means that the observables $\Xm_{j,p,j+2k,p}$ are purely real, and observables $\Xm_{(j,p),(j+2k+1,p)}$ are purely imaginary. The observables on the rungs satisfy $\Xm_{(j,p),(j,3-p)} = -\Xm_{(j,p),(j,3-p)}^{\ast}$, therefore they are purely imaginary which implies that the kinetic energy on the rungs is $0$. 

\end{appendix}


\end{document}